\documentclass{aa} 

\bibpunct{(}{)}{;}{a}{}{,} 

\usepackage[varg]{txfonts}
\usepackage{amssymb}
\usepackage{amsmath}
\usepackage{natbib}
\usepackage{subcaption}
\usepackage{graphicx}
\usepackage{epstopdf}
\usepackage{marvosym}
\usepackage{epsfig}
\usepackage{multirow}
\usepackage{tikz}
\usepackage{nicefrac}
\usepackage{arydshln}
\usepackage{url}
\usepackage{blkarray}
\usepackage{hyphenat} 
\usepackage{lipsum}
\usepackage{caption}
\usepackage{booktabs}
\usetikzlibrary{arrows,calc,intersections,positioning,arrows,decorations.pathmorphing,decorations.markings,shapes}

\newcommand{\RN}[1]{
  \textup{\uppercase\expandafter{\romannumeral#1}}
}

\begin{document}

\title{Spacecraft VLBI tracking to enhance stellar occultations astrometry of planetary satellites}

\author{M. Fayolle \inst{\ref{TUDELFT}}
        \and V. Lainey \inst{\ref{IMCCE}} \and D. Dirkx \inst{\ref{TUDELFT}} \and L.I. Gurvits \inst{\ref{JIVE},\ref{TUDELFT}} \and G. Cimo \inst{\ref{JIVE}} \and S. J. Bolton \inst{\ref{SWRI}}}

\institute{Delft University of Technology, Kluyverweg 1, 2629HS Delft, The Netherlands \\ \email{m.s.fayolle-chambe@tudelft.nl} \label{TUDELFT}
\and IMCCE, Observatoire de Paris, 77 Av. Denfert-Rochereau, 75014, Paris, France \label{IMCCE}
\and Joint Institute for VLBI ERIC, Oude Hoogeveensedijk 4, 7991PD, Dwingeloo, the Netherlands\label{JIVE}
\and Southwest Research Institute, San Antonio, TX, USA \label{SWRI}}

\date{Received 26 Mai 2023 / Accepted 12 July 2023}

\abstract
{Stellar occultations currently provide the most accurate ground-based measurements of the positions of natural satellites  (down to a few kilometres for the Galilean moons). However, when using these observations in the calculation of satellite ephemerides, the uncertainty in the planetary ephemerides dominates the error budget of the occultation.}
{We quantify the local refinement in the central planet's position achievable by performing Very Long Baseline Interferometry (VLBI) tracking of an in-system spacecraft temporally close to an occultation. We demonstrate the potential of using VLBI to enhance the science return of stellar occultations for satellite ephemerides.} 
{We identified the most promising observation and tracking opportunities offered by the Juno spacecraft around Jupiter as perfect test cases, for which we ran simulations of our VLBI experiment.} 
{VLBI tracking at Juno's perijove close to a stellar occultation locally (in time) reduces the uncertainty in Jupiter's angular position in the sky to 250-400 m. This represents up to an order of magnitude improvement with respect to current solutions and is lower than the stellar occultation error, thus allowing the moon ephemeris solution to fully benefit from the observation.}
{Our simulations showed that the proposed tracking and observation experiment can efficiently use synergies between ground- and space-based observations to enhance the science return on both ends. The reduced error budget for stellar occultations indeed helps to improve the moons' ephemerides, which in turn benefit planetary missions and their science products, such as the recently launched JUICE and upcoming Europa Clipper missions.} 
   

\maketitle

\section{Introduction} \label{sec:introduction}

In addition to classical astrometry, various ground-based observation techniques have been developed and intensively used to study the orbital motion of natural satellites \citep[e.g.][and references therein]{arlot2019}. Observations of stellar occultations, which occur when a moon passes in front a star, have proven particularly promising \citep{morgado2019,morgado2022}. Thanks to the Gaia star catalogues, with sub-mas (milliarcsecond) precision for the star  positions \citep{gaia2018,brown2021}, stellar occultations provide the most accurate ground-based measurements to date for natural satellite positions \citep[accuracy of the order of 1 mas, i.e. a few km for the Galilean satellites,][]{morgado2022}. 

\begin{figure}[tbp!]
        \centering
        \begin{minipage}[l]{1.0\columnwidth}
                \centering
                \makebox[\textwidth][c]{\includegraphics[width=1.0\textwidth]{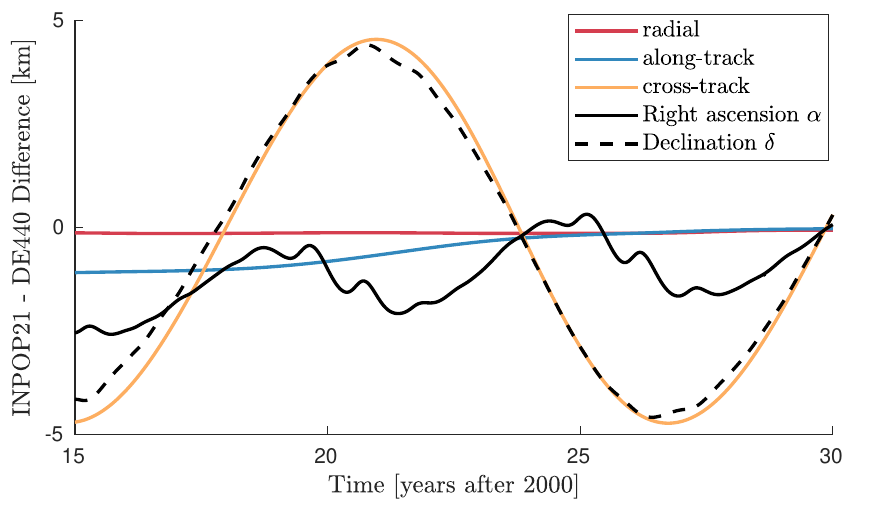}}
        \end{minipage}
        \caption{Difference between INPOP21 and DE440 planetary ephemerides for Jupiter in radial, along-track (tangential), and cross-track (out-of-plane) directions. Deviations in right ascension $\alpha$ and declination $\delta$ are also provided.}
        \label{fig:ephemeridesComparison}
\end{figure}

These observations constrain the moons' positions in the plane of the sky, typically in the International Celestial Reference Frame (ICRF). However, improving satellite ephemerides requires information on the moons' relative positions with respect to the central planet, rather than their absolute positions in the ICRF. To use stellar occultations in satellite ephemeris generation, the uncertainty in the planet's position thus directly increases the effective error budget of the stellar occultations. 

For recent occultations by the Galilean satellites, discrepancies between observed and predicted events (the latter being ephemerides-based) are still significant and vary depending on which planetary ephemerides are considered, as reported in \cite{morgado2022}. Non-negligible differences indeed remain between different Jovian ephemerides, indicating possible errors or discrepancies. This is illustrated in Fig. \ref{fig:ephemeridesComparison} for the most recent solutions: INPOP21 \citep{fienga2021} and DE440 \citep{park2021}. The deviations are small for the in-plane components, especially in the radial direction, which significantly benefited from Juno tracking data \citep[e.g.][]{fienga2019}. The discrepancy is  larger, however,  in the out-of-plane direction, with a long-term periodic effect building up to 4.5 km. In right ascension and declination, differences can amount up to 2 km and 5 km, respectively, which is comparable to a typical stellar occultation accuracy.

A possible means to mitigate this error source is to combine spacecraft VLBI tracking with stellar occultation observations. To demonstrate the added value of such an experiment, we quantify the local refinement in Jupiter’s position provided by phase-referencing VLBI observations of an orbiting spacecraft in the close vicinity of stellar occultations. Phase-referenced VLBI tracking relies on a nearby radio source (within a few degrees of the spacecraft) to perform phase calibration and obtain accurate measurements of a spacecraft’s position in the ICRF \citep[e.g.][]{jones2010,duev2012,duev2016}. If the spacecraft orbits close to Jupiter, this also provides valuable constraints on Jupiter's position in the ICRF. It is worth mentioning that we focus on Jupiter's angular position ($\alpha_\mathrm{Jup},\delta_\mathrm{Jupiter}$) in the sky, which directly affects the stellar occultation error budget (Section \ref{sec:experimentPrinciple}), and do not intend to improve Jupiter's global fit. We exploit the presence of the Juno spacecraft in the Jovian system \citep{bolton2017} and use two experiment opportunities that it offers, in 2023 and 2024, as test cases for our study. 

To quantify the improvement in the effective stellar occultation error budget provided by the VLBI data, we can use the INPOP21-DE440 deviations (Fig. \ref{fig:ephemeridesComparison}) as a conservative lower limit for the current uncertainty in Jupiter's position. Our two test occultations of interest, in 2023 and 2024, coincide with Jupiter's crossing the ecliptic. It thus also corresponds to a local minimum for the INPOP21-DE440 difference, mostly originating from a small discrepancy in Jupiter's orbit orientation. Consequently, the periodic behaviour observed in Fig. \ref{fig:ephemeridesComparison} illustrates the discrepancies between the two fits, but is likely not indicative of the exact evolution of the ephemeris error in time. This would imply that the uncertainty in Jupiter's out-of-plane position at time $t$ can in practice be expected to take any value up to $\sim$4.5 km. We thus chose the averaged difference between INPOP21 and DE440 as a better metric for Jupiter's position error.  

The principle of the experiment is described in Section \ref{sec:experimentAndSimulations}, where upcoming tracking and observation opportunities are also identified to be used as test cases. Section \ref{sec:experimentAndSimulations} presents the simulations performed for two stellar occultations, to demonstrate the local improvement in Jupiter's right ascension and declination accuracy, and the resulting improvement in stellar occultation quality for satellite ephemerides. The results and conclusions are discussed in Sections \ref{sec:results} and \ref{sec:conclusions}, respectively.

\section{Experiment principle and simulations} \label{sec:experimentAndSimulations}

\subsection{Experiment and next opportunities} \label{sec:experimentPrinciple}

\begin{figure}[tbp!]
        \centering
        \begin{minipage}[l]{1.0\columnwidth}
                \centering
                \makebox[\textwidth][c]{\includegraphics[width=1.0\textwidth]{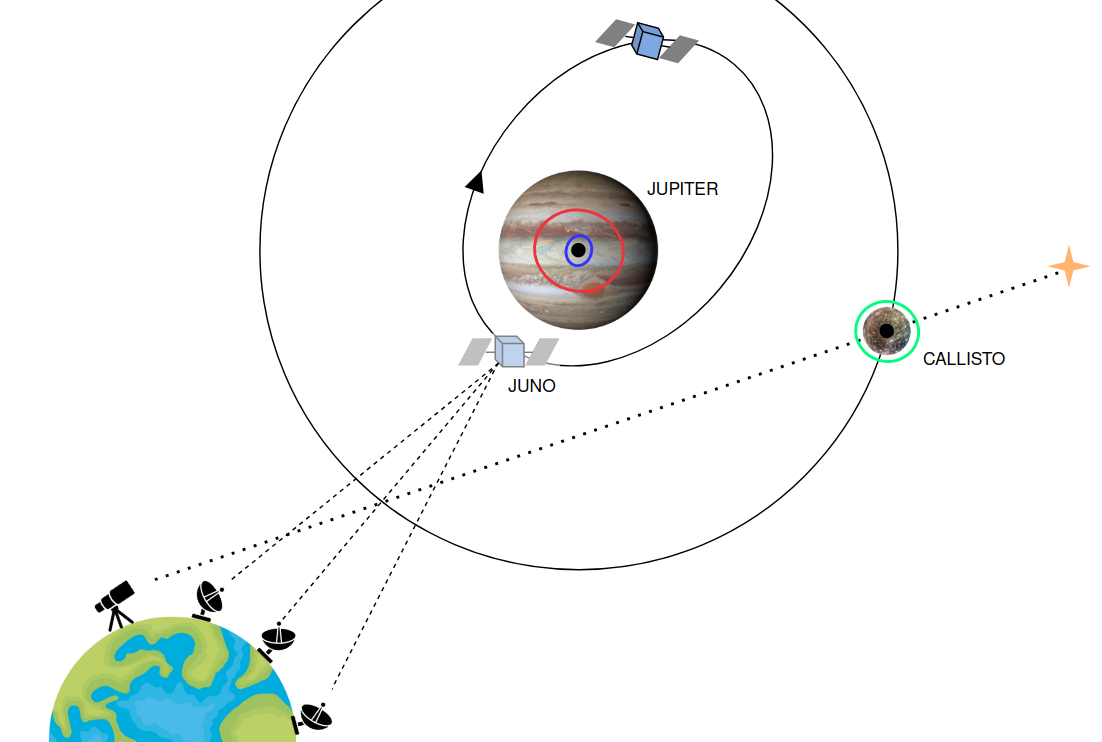}}
        \end{minipage}
        \caption{Illustration of the proposed experiment (not to scale). The occultation yields a very accurate measurement of Callisto's position in the ICRF (small green ellipse centred at Callisto). Tracking the Juno spacecraft at the perijove(s) closest to the occultation would reduce Jupiter's initial position uncertainty (red ellipse) to the smaller  blue ellipse.}
        \label{fig:experiment}
\end{figure}

Figure \ref{fig:experiment} summarises the configuration of the proposed experiment. A stellar occultation by Callisto is used as an example and would nominally measure the moon's position ($\alpha$, $\delta$) in the ICRF to an accuracy of a few kilometres (green ellipse). Reconstructing the moon's orbit around Jupiter requires  accounting for Jupiter's own position error (assuming the two errors are uncorrelated) as
\begin{align}
\sigma^2\left(\boldsymbol{r}_\mathrm{Callisto/Jup}\right) = \sigma^2\left(\boldsymbol{r}_\mathrm{Callisto}\right) + \sigma^2\left(\boldsymbol{r}_\mathrm{Jup}\right),     
\end{align}
where $\boldsymbol{r}_\mathrm{Callisto/Jup}$ denotes Callisto's position with respect to Jupiter, while $\boldsymbol{r}_\mathrm{Callisto}$ is the moon's position with respect to the Solar System barycentre (SSB), as provided by the stellar occultation. However, Jupiter's ephemeris error is similar to or possibly larger than the occultation uncertainty (red ellipse in Fig. \ref{fig:experiment}). VLBI tracking of Juno during the perijove(s) closest to the occultation would help refine Jupiter's barycentric position, as was already done in the past by \cite{jones2019,jones2021}. 

Since each of Juno's orbits lasts about 40 days, the occultation might occur a few weeks away from the closest perijove. We therefore propose to track the spacecraft during the two perijoves surrounding the stellar occultation. This would constrain Jupiter's position both before and after the observation, thus ensuring a reduced uncertainty at occultation time.

\begin{table*}
        \small
        \caption{Selected phase calibrators for each of the four perijoves.}
        \label{tab:calibrators}
        \centering
        \begin{tabular}{ c c c c c c}
                \hline
                \textbf{Juno perijoves} & \textbf{Name calibrator} & \textbf{Separation with target} & \multicolumn{2}{c}{\textbf{Position uncertainty}} & \textbf{Total flux density}\\
                & & [deg] & $\sigma(\alpha)$ [nrad] & $\sigma(\delta)$ [nrad] & [Jy] \\ \hline
                15-10-2023 & J0244+1320 & 1.10 & 0.5 & 1.0 & 0.207 \\
                22-11-2023 & J0225+1134 & 1.37 & 0.7 & 0.8 & 0.313 \\ \hline
                30-12-2023 & J0211+1051 & 1.40 & 0.5 & 0.8 & 0.516 \\
                03-02-2024 & J0225+1134 & 1.81 & 0.7 & 0.8 & 0.313 \\ \hline
        \end{tabular}
\end{table*}

\begin{table*}[tpb!]
        \small
        \caption{Predicted occultations by the Galilean satellites  and corresponding Juno perijoves. }
        \label{tab:predictions}
        \centering
        \begin{tabular}{c c | c  c |  c c c}
                \hline
                \multicolumn{2}{c|}{\textbf{Occultations}} & \multicolumn{2}{c|}{\textbf{Juno perijoves}}  & \multicolumn{3}{c}{\textbf{VLBI networks coverage}} \\
                Date & Occulting moon & Date & Time [UTC] & EVN & VLBA & LBA  \\ \hline
                \textbf{23-10-2023} & \textbf{Ganymede} & 15-10-2023 & 10:52:58 & 0 \% & 100 \% & 55 \% \\
                & & 22-11-2023 & 12:16:47 & 0 \% & 45 \% & 100 \% \\ \hline
                \textbf{15-01-2024} & \textbf{Callisto} & 30-12-2023 & 12:36:20 & 42.8 \% & 0 \% & 100 \% \\
                & & 03-02-2024 & 21:47:30 & 80 \% & 100 \% & 0 \% \\ \hline
        \end{tabular}
        \tablefoot{Stellar occultations were predicted for the years 2023 and 2024 only. For each VLBI network, the coverage percentage indicates the fraction of the 6h arc during which more than three stations (per network) can track Juno's radio signal (minimum elevation of 10 deg).}
\end{table*}

Two promising occultations will occur in the near future, one by Ganymede on 23 October 2023 (star magnitude $G = 11.3$) and one by Callisto on 15 January 2024 ($G=8.8$). Table \ref{tab:predictions} provides the dates and times of the Juno perijoves preceding and following these two occultations. For each of these perijoves we identified suitable phase calibrators within two degrees of the Juno spacecraft. The sources taken into consideration for this work are listed in Table \ref{tab:calibrators}. We also ran a coverage analysis for Juno tracking from three major VLBI telescope networks: the European VLBI Network\footnote{\url{https://www.evlbi.org/}} (EVN), the  Very Long
Baseline Array\footnote{\url{https://public.nrao.edu/telescopes/vlba/}} (VLBA), and the Long Baseline Array\footnote{\url{https://www.atnf.csiro.au/vlbi/overview/index.html}} (LBA). None of the networks can alone ensure tracking during the four perijoves of interest. In particular, EVN cannot cover the perijoves surrounding Ganymede's 2023 occultation, while VLBA and LBA each miss one perijove of Callisto's 2024 occultation.

\subsection{Simulation set-up} \label{sec:simulationSetup}

\begin{figure*}[tbp!]
        \centering
        \begin{minipage}[c]{1.0\textwidth}
                \centering
                \makebox[\textwidth][c]{\includegraphics[width=1.1\textwidth]{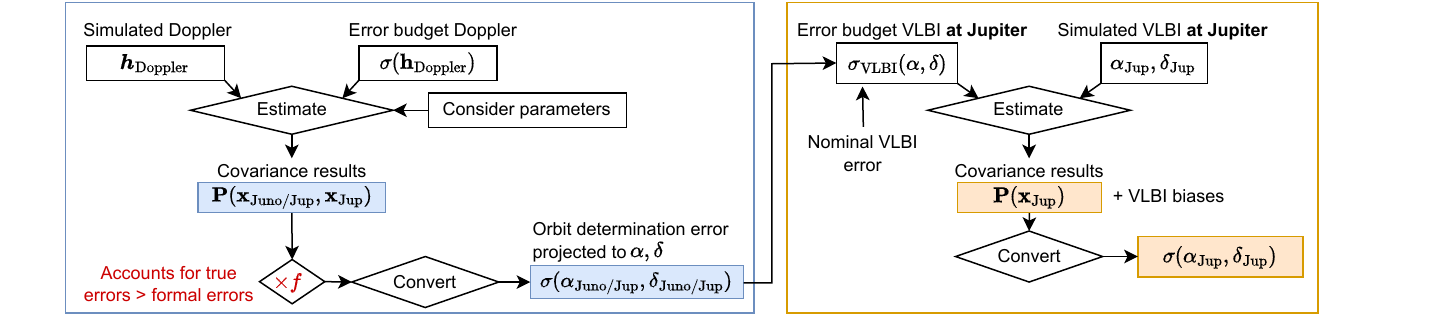}}
        \end{minipage}
        \caption{Workflow used to quantify the reduction in Jupiter's position uncertainty (in the plane of the sky) achievable with the proposed VLBI experiment.}
        \label{fig:flowchart}
\end{figure*}

The aim of our analysis is to quantify the local (in time) uncertainty reduction in Jupiter's right ascension and declination at the time of the occultation(s) using VLBI. As illustrated in Fig. \ref{fig:flowchart}, this was achieved in two steps. First, we determined the error associated with Juno's orbit (Fig. \ref{fig:flowchart}, left), referred to as Juno state estimation. Second, we used Juno's estimated orbit uncertainty to construct VLBI observables re-centred at Jupiter with realistic errors. We could subsequently estimate Jupiter's state at the time of the occultation from these VLBI data points (Fig. \ref{fig:flowchart}, right) referred to as Jupiter state estimation.

It should be noted that this study relied on simulated observations and was limited to covariance analyses. This approach is well adapted to quantify the contribution of VLBI measurements to Jupiter's local position, even if the real data analysis will later require a full fit. All results, from both  Juno's and Jupiter state estimations, are therefore based on formal uncertainties derived from covariance matrices. 

Starting with Juno state estimation, we simulated Doppler measurements during the perijoves preceding and following each of the two occultation opportunities (Section \ref{sec:experimentPrinciple}). Doppler data were generated every 60s over 6h tracking arcs, with a noise of 0.05 mm/s in agreement with the residuals from Juno radio-science experiment \citep{iess2018}.

For the purpose of our analysis, which focused on a local rather than global fit improvement for Jupiter, we only needed to consider two perijoves for each occultation. As a consequence, we chose not to estimate all dynamical parameters usually determined from Juno data. Only Juno's and Jupiter's states were estimated at perijove time $t_\mathrm{PJ}$, the latter ensuring that the Jovian ephemeris uncertainty was included in Juno's orbit error. Similarly, we also added a number of consider parameters to account for their influence on the estimation \citep[e.g.][]{montenbruck2002}:
\begin{itemize}
        \item Jupiter's spherical harmonics gravity coefficients up to degree and order 2, and zonal coefficients up to degree 10;
        \item Jupiter's pole orientation and rotation rate;
        \item Empirical accelerations on Juno, required to fit Doppler data. We assumed constant components in the  radial, tangential, and normal (RTN) directions, estimated every 10 min during a two-hour window around perijove time, as described in \cite{durante2020}.
\end{itemize}
The uncertainty values for all the consider parameters were taken from the estimation results at mid-Juno mission \citep{iess2018,durante2020} and are reported in Table \ref{tab:apriori}.

Doppler data alone cannot notably improve Jupiter's state uncertainties beyond their a priori constraints. The main objective of this first estimation step, however, is to obtain the covariance describing Juno's orbit uncertainty $\mathbf{P}(\mathbf{x_\mathrm{Juno/Jup}})(t_\mathrm{PJ})$, which can be extracted from the full covariance matrix and directly used to recentre VLBI observables to Jupiter's centre of mass (Fig. \ref{fig:flowchart}), using the same methodology as in \cite{dirkx2017}. The total uncertainty $\sigma_\mathrm{VLBI}(\alpha, \delta)$ for these observables needs to account for the nominal error budget for VLBI observations $\sigma_\star(\alpha,\delta)$, and for Juno's orbit error
\begin{align}
\sigma^2_\mathrm{VLBI}(\alpha,\delta)(t) = \sigma^2_\star(\alpha,\delta) +  \sigma^2(\mathbf{\alpha_\mathrm{Juno/Jup},\delta_\mathrm{Juno/Jup}})(t), \label{eq:errorModel}
\end{align}
where $\sigma(\mathbf{\alpha_\mathrm{Juno/Jup},\delta_\mathrm{Juno/Jup}})(t)$ is the projection of the propagated covariance $\mathbf{P}(\mathbf{x_\mathrm{Juno/Jup}})(t)$ to right ascension and declination. The noise value of a single VLBI observation $\sigma_\star(\alpha,\delta)$ was conservatively set to 1.0 nrad based on recent phase-referencing VLBI tracking of Cassini and Mars Express spacecraft \citep{jones2010,jones2019,duev2016}. Since formal uncertainties are typically known to be too optimistic compared to the true errors \citep[e.g.][]{dirkx2017}, a factor $f$ was first applied to Juno's state covariance before propagating it from perijove to occultation time. We used both $f=1$ and $f=5$, but our nominal results, unless otherwise indicated, correspond to the latter conservative case to ensure that Juno's orbit error is not underestimated. 

Independent VLBI observables re-centred to Jupiter were simulated every 20 minutes, using the error model in Eq. \ref{eq:errorModel}. A systematic bias was also added to these observations, corresponding to the uncertainty in the phase calibrator's position in the ICRF (Table \ref{tab:calibrators}). These biases were included as consider parameters. As such, they cannot be reduced in the estimation process, which ensures that this uncertainty source is conservatively accounted for in our results. Four additional tracking configurations were considered: tracking by the EVN, VLBA, and LBA networks individually, as well as a perfect coverage case where all three are involved. From the VLBI data, we estimated Jupiter's state at occultation time. The resulting uncertainties in Jupiter's right ascension and declination $\sigma(\alpha_\mathrm{Jup}, \delta_\mathrm{Jup})$ correspond to the blue error ellipse in Fig. \ref{fig:experiment} and are   discussed in the following section.

\section{Expected contribution} \label{sec:results}

\begin{figure}[tbp!]
        \centering
        \begin{minipage}[c]{1.0\columnwidth}
                \centering
                \makebox[\columnwidth][c]{\includegraphics[width=1.0\columnwidth]{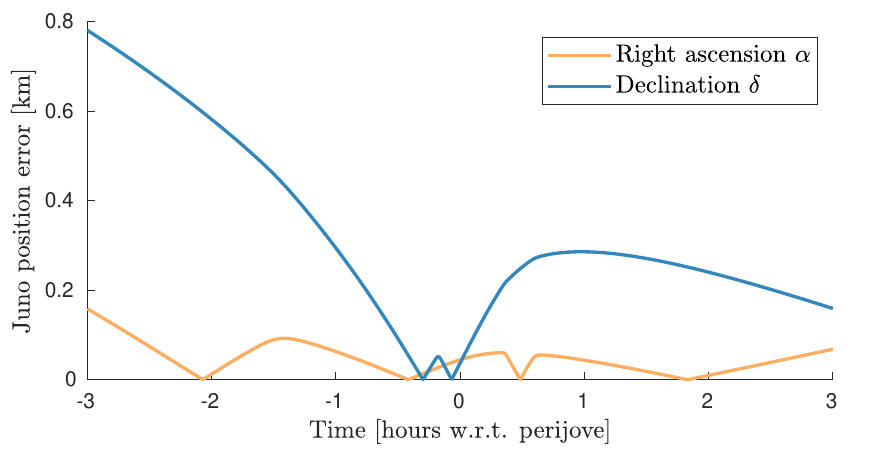}}
        \end{minipage}
        \caption{Propagated errors in Juno's right ascension and declination for the 15 October 2023 perijove, preceding Ganymede's occultation. The results correspond to the $f=1$ case (i.e. no scaling of Juno's determination orbit error).}
        \label{fig:propagatedErrorsJuno}
\end{figure}

\begin{table*}[tpb!]
        \small
        \caption{Formal errors in Jupiter's position.}
        \label{tab:results}
        \centering
        \begin{tabular}{c | c c | c c | c c | c c}
                \hline
                \multicolumn{1}{c}{\textbf{VLBI network(s)}} & \multicolumn{4}{c}{\textbf{23-10-2023 occultation}} & \multicolumn{4}{c}{\textbf{15-01-2024 occultation}}  \\ 
                & \multicolumn{2}{c|}{$\sigma(\alpha_\mathrm{Jup})$ [km]} & \multicolumn{2}{c|}{$\sigma(\delta_\mathrm{Jup})$ [km]} &  \multicolumn{2}{c|}{$\sigma(\alpha_\mathrm{Jup})$ [km]}& \multicolumn{2}{c}{$\sigma(\delta_\mathrm{Jup})$ [km]} \\
                & $f=1$ & $f=5$ & $f=1$ & $f=5$ & $f=1$& $f=5$& $f=1$ & $f=5$ \\\hline
                EVN & N.A. & N.A. & N.A. & N.A. & 0.32 & 0.36 & 0.49 & 0.80 \\
                VLBA & 0.28 & 0.31 & 0.54 & 0.63 & 0.43 & 0.45 & 0.81 & 0.92 \\
                LBA & 0.30 & 0.31 & 0.51 & 0.58 & 0.38 & 0.41 & 0.73 & 0.85 \\ \hline
                All & 0.28 & 0.29 & 0.51 & 0.58 & 0.30 & 0.32 & 0.46 & 0.60 \\ \hline
        \end{tabular}
        \tablefoot{The formal uncertainties are provided for different VLBI tracking configurations, and are expressed as uncertainties in right ascension and declination at the occultation time. Results are given for $f=1$ and $f=5$, $f$ being the factor applied to Juno's state covariance to re-scale the orbit error.}
\end{table*}

\begin{figure*} [tbp!] 
        \centering
        \begin{minipage}[l]{1.0\columnwidth}
                \centering
                \subcaptionbox{\label{fig:occultation_2310}}
                {\includegraphics[width=1.1\textwidth]{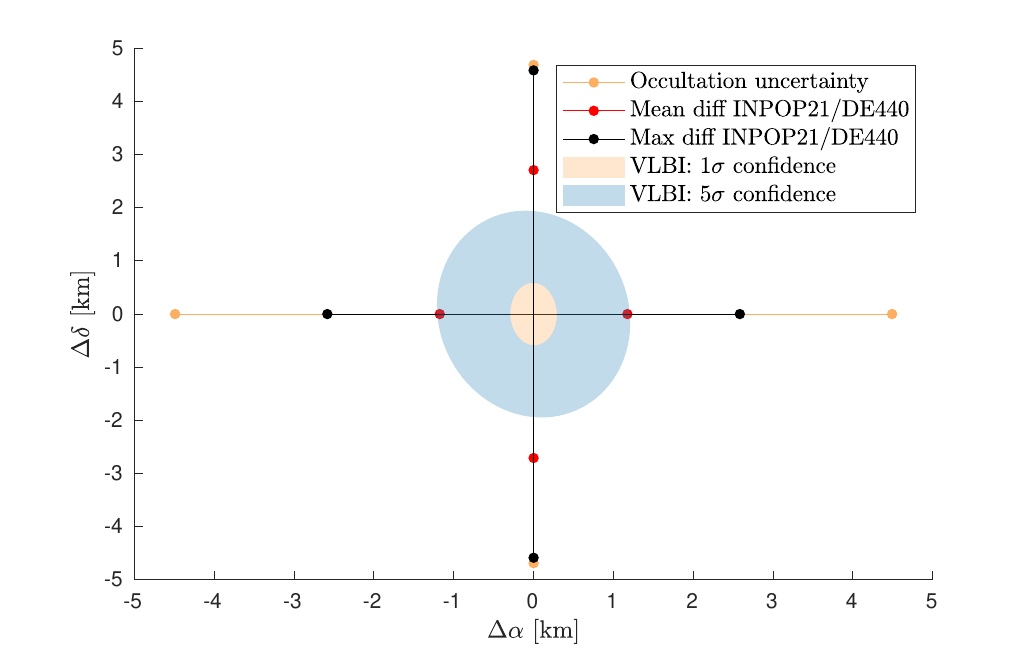}}
        \end{minipage}
        \hfill{} 
        \begin{minipage}[r]{1.0\columnwidth}
                \centering
                \subcaptionbox{\label{fig:occultation_1501}}
                {\includegraphics[width=1.1\textwidth]{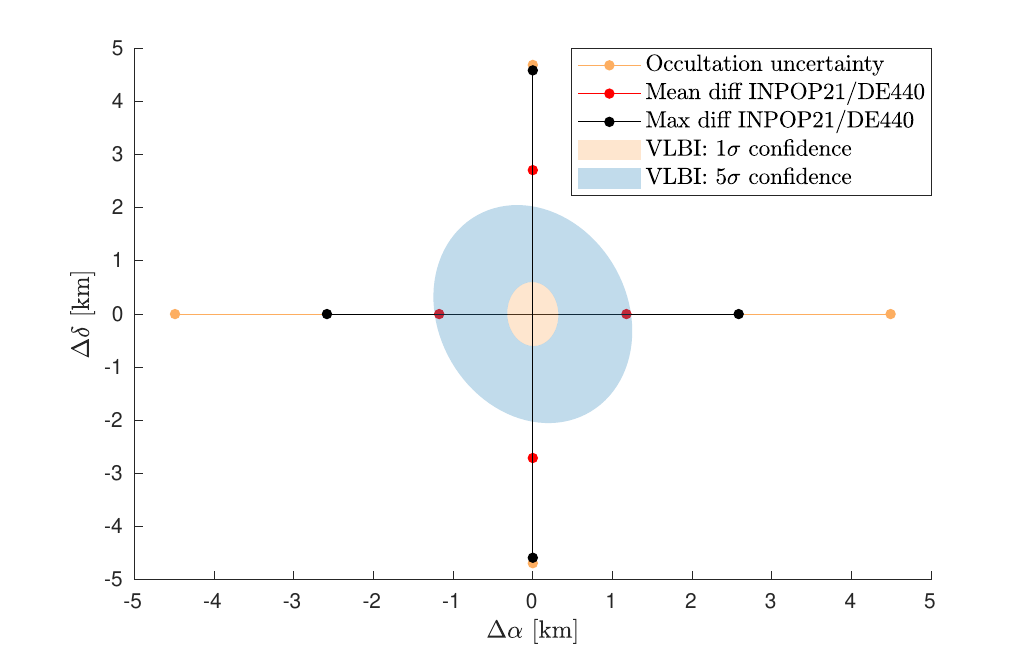}}
        \end{minipage}
        \caption{Uncertainties in $\alpha_\mathrm{Jup}$ and $\delta_\mathrm{Jup}$ at occultation time. Panel a: Occultation by Ganymede on 23 October 2023; Panel b: occultation by Callisto on 15 January 2024. Markers on the x- and y-axes indicate the averaged and maximum deviation between INPOP21 and DE440 (over the period 2015--2030), as well as the typical uncertainty for stellar occultations \citep[based on][]{morgado2022}. The coloured confidence ellipses represent the $1\sigma$ (orange) and $5\sigma$ (blue) covariances in Jupiter's position resulting from VLBI tracking.}
        \label{fig:occultations}
\end{figure*}

From the simulated Doppler measurements, we first estimated Juno's state with respect to Jupiter at perijove time. We obtained formal uncertainties in right ascension  and declination $\sigma(\alpha_\mathrm{Juno/Jup},\delta_\mathrm{Juno/Jup})$ between 50 and 120 m (perijove-dependent). Those uncertainties and their correlations were then propagated over the entire arc, as shown in Fig. \ref{fig:propagatedErrorsJuno} for the 15 October 2023 perijove. Jupiter-centred VLBI measurements could then be constructed at any time $t$ from the instantaneous orbit determination error. 

Uncertainties in $\alpha_\mathrm{Jup},\delta_\mathrm{Jup}$ estimated from the VLBI observations are displayed in Fig. \ref{fig:occultations}. The orange and blue ellipses represent the $1\sigma$ and $5\sigma$ covariances (in both cases a factor $f=5$ was first applied to Juno's orbit error). The latter is a very conservative case, again accounting for formal errors possibly being too optimistic. For both occultations, VLBI tracking leads to $1\sigma$ errors of 300 m and 600 m, for $\alpha_\mathrm{Jup}$ and $\delta_\mathrm{Jup}$ respectively (in the worst-case scenario, i.e. $f=5$). This is well below both the stellar occultations accuracy and the estimated error of the current Jupiter ephemeris, and would thus allow the moons' ephemerides to fully benefit from the exceptional quality of these observations.

The estimated errors in $\alpha_\mathrm{Jup}$ and $\delta_\mathrm{Jup}$ are respectively about a factor of 4 and a factor of 5 smaller than the average difference between the two ephemerides solutions. With respect to the maximum INPOP21-DE440 deviation, the uncertainty reduction almost reaches a factor of  10. Even when considering the very pessimistic 5$\sigma$ confidence ellipse, a significant improvement is still attainable for $\sigma(\delta_\mathrm{Jup})$. 

As mentioned in Section \ref{sec:introduction}, the differences between the current ephemerides give a conservative estimate of Jupiter's state uncertainty. The two ephemerides are based on the same observation set \citep{fienga2021,park2021} and rely on comparable dynamical models, and may therefore possess common biases. VLBI tracking, on the other hand, provides the absolute measures of Jupiter's position in the sky, with biases at the sub-nrad level. The local improvement in the Jovian ephemeris provided by VLBI may thus be greater than our results indicate. 

Finally, the results shown in Fig. \ref{fig:occultations} assumed continuous VLBI tracking during the 6h arcs, which would require several networks to be involved  (Table \ref{tab:predictions}). Table \ref{tab:results} presents the outcome of several tracking scenarios. For Ganymede's occultation on 23 October 2023, only relying on either VLBA or LBA is sufficient to ensure errors comparable to those obtained with the two networks (irrespective of the factor $f$ applied to Juno's orbit error). This does not hold for Callisto's occultation  which would benefit from using multiple networks, especially in the $f=5$ case. To optimise the outcome of the experiment, relying on two or three VLBI networks for each perijove would thus be ideal.

\section{Conclusions} \label{sec:conclusions}

To optimise the science return of stellar occultations for satellite ephemerides calculations, VLBI tracking of an in-system spacecraft can be used to locally reduce the uncertainty in the central planet’s position, which directly contributes to the occultation error budget. To demonstrate the potential of this VLBI experiment, we performed simulations for two promising observation opportunities with the Juno spacecraft, in 2023 and 2024. For both test cases our results indicate that VLBI tracking will indeed reduce the uncertainty in Jupiter's position to the sub-kilometre level at occultation time (Section \ref{sec:results}), ensuring that it no longer dominates the stellar occultation error budget.

This also represents a unique opportunity to test our current planetary and satellites ephemerides, which are both involved in the prediction of stellar occultations, and both have estimated errors at a level similar to or higher than the stellar occultations. In practice, offsets between predicted and observed occultations already indicated possible errors and/or discrepancies in the existing solutions \citep{morgado2022}. Our experiment could help quantify them, possibly identifying their origin and distinguishing between planetary and satellite ephemerides errors. 

Finally, this experiment would serve as a preparation for the upcoming JUICE and Europa Clipper missions. First, it would help to improve the Galilean satellites' ephemerides before the missions, which can reduce pre- and post-flyby corrective manoeuvres \citep{bellerose2016,boone2017,hener2022}. Moreover, if proven successful, similar experiments could be implemented for any other mission, including JUICE and Clipper. By exploiting synergies between different measurement techniques, which will likely be critical in order to achieve a high-accuracy ephemeris solution from the missions' data \citep{fayolle2022}, it would capitalise on the presence of one or more in-system spacecraft to also benefit ground-based observations, and therefore enhance the science return of the mission(s). Among other advantages, radio occultation studies could  benefit from the VLBI tracking experiments proposed here, the Doppler data being directly useful for such analyses \citep{bocanegra2019}, while VLBI measurements can refine Jupiter's local state at occultation time.

\begin{acknowledgements} 
We would like to thank Josselin Desmars for providing us with stellar occultations predictions for the coming years.
This research was partially funded by ESA's OSIP (Open Space Innovation Platform)
program, and supported by CNES, focused on PRIDE and JUICE. We are grateful to the anonymous referee for their very useful comments.
\end{acknowledgements} 

\bibliographystyle{aa} 
\bibliography{references}

\begin{thebibliography}{23}
\expandafter\ifx\csname natexlab\endcsname\relax\def\natexlab#1{#1}\fi

\bibitem[{Arlot(2019)}]{arlot2019}
Arlot, J. 2019, Journal of Astronomical History and Heritage, 22, 78

\bibitem[{Bellerose {et~al.}(2016)Bellerose, Nandi, Roth, Tarzi, Boone,
  Criddle, \& Ionasescu}]{bellerose2016}
Bellerose, J., Nandi, S., Roth, D., {et~al.} 2016

\bibitem[{Bocanegra-Baham{\'o}n {et~al.}(2019)Bocanegra-Baham{\'o}n,
  Calv{\'e}s, Gurvits, Cim{\`o}, Dirkx, Duev, Pogrebenko, Rosenblatt, Limaye,
  Cui, {et~al.}}]{bocanegra2019}
Bocanegra-Baham{\'o}n, T., Calv{\'e}s, G.~M., Gurvits, L., {et~al.} 2019,
  Astronomy \& Astrophysics, 624, A59

\bibitem[{Bolton {et~al.}(2017)Bolton, Lunine, Stevenson, Connerney, Levin,
  Owen, Bagenal, Gautier, Ingersoll, Orton, {et~al.}}]{bolton2017}
Bolton, S., Lunine, J., Stevenson, D., {et~al.} 2017, Space Science Reviews,
  213, 5

\bibitem[{Boone {et~al.}(2017)Boone, Bellerose, \& Roth}]{boone2017}
Boone, D., Bellerose, J., \& Roth, D. 2017

\bibitem[{Brown {et~al.}(2021)Brown, Vallenari, Prusti, De~Bruijne, Babusiaux,
  Biermann, Creevey, Evans, Eyer, Hutton, {et~al.}}]{brown2021}
Brown, A.~G., Vallenari, A., Prusti, T., {et~al.} 2021, Astronomy \&
  Astrophysics, 649, A1

\bibitem[{Dirkx {et~al.}(2017)Dirkx, Gurvits, Lainey, Lari, Milani, Cim{\`o},
  Bocanegra-Bahamon, \& Visser}]{dirkx2017}
Dirkx, D., Gurvits, L.~I., Lainey, V., {et~al.} 2017, Planetary and Space
  Science, 147, 14

\bibitem[{Duev {et~al.}(2012)Duev, Molera~Calv{\'e}s, Pogrebenko, Gurvits,
  Cimo, \& Bahamon}]{duev2012}
Duev, D.~A., Molera~Calv{\'e}s, G., Pogrebenko, S.~V., {et~al.} 2012, Astronomy
  \& Astrophysics, 541, A43

\bibitem[{Duev {et~al.}(2016)Duev, Pogrebenko, Cim{\`o}, Calv{\'e}s,
  Baham{\'o}n, Gurvits, Kettenis, Kania, Tudose, Rosenblatt,
  {et~al.}}]{duev2016}
Duev, D.~A., Pogrebenko, S.~V., Cim{\`o}, G., {et~al.} 2016, Astronomy \&
  Astrophysics, 593, A34

\bibitem[{Durante {et~al.}(2020)Durante, Parisi, Serra, Zannoni, Notaro,
  Racioppa, Buccino, Lari, Gomez~Casajus, Iess, {et~al.}}]{durante2020}
Durante, D., Parisi, M., Serra, D., {et~al.} 2020, Geophysical Research
  Letters, 47, e2019GL086572

\bibitem[{Fayolle {et~al.}(2022)Fayolle, Dirkx, Lainey, Gurvits, \&
  Visser}]{fayolle2022}
Fayolle, M., Dirkx, D., Lainey, V., Gurvits, L., \& Visser, P. 2022, Planetary
  and Space Science, 219, 105531

\bibitem[{Fienga {et~al.}(2021)Fienga, Deram, Di~Ruscio, Viswanathan, Camargo,
  Bernus, Gastineau, \& Laskar}]{fienga2021}
Fienga, A., Deram, P., Di~Ruscio, A., {et~al.} 2021, INPOP21a planetary
  ephemerides

\bibitem[{Fienga {et~al.}(2019)Fienga, Deram, Viswanathan, Di~Ruscio, Bernus,
  Durante, Gastineau, \& Laskar}]{fienga2019}
Fienga, A., Deram, P., Viswanathan, V., {et~al.} 2019, INPOP19a planetary
  ephemerides

\bibitem[{Gaia {et~al.}(2018)Gaia, Brown, Vallenari, Prusti, De~Bruijne,
  Babusiaux, Juh{\'a}sz, Marschalk{\'o}, Marton, Moln{\'a}r,
  {et~al.}}]{gaia2018}
Gaia, C., Brown, A., Vallenari, A., {et~al.} 2018, Astronomy \& Astrophysics,
  616

\bibitem[{Hener(2022)}]{hener2022}
Hener, J. 2022

\bibitem[{Iess {et~al.}(2018)Iess, Folkner, Durante, Parisi, Kaspi, Galanti,
  Guillot, Hubbard, Stevenson, Anderson, {et~al.}}]{iess2018}
Iess, L., Folkner, W., Durante, D., {et~al.} 2018, Nature, 555, 220

\bibitem[{Jones {et~al.}(2021)Jones, Folkner, Park, \& Dhawan}]{jones2021}
Jones, D., Folkner, W., Park, R., \& Dhawan, V. 2021in , 142--05

\bibitem[{Jones {et~al.}(2010)Jones, Fomalont, Dhawan, Romney, Folkner, Lanyi,
  Border, \& Jacobson}]{jones2010}
Jones, D.~L., Fomalont, E., Dhawan, V., {et~al.} 2010, The Astronomical
  Journal, 141, 29

\bibitem[{Jones {et~al.}(2019)Jones, Romney, Folkner, Park, Jacobs, \&
  Dhawan}]{jones2019}
Jones, D.~L., Romney, J.~D., Folkner, W.~M., {et~al.} 2019, in 2019 IEEE
  Aerospace Conference, IEEE, 1--6

\bibitem[{Montenbruck {et~al.}(2002)Montenbruck, Gill, \&
  Lutze}]{montenbruck2002}
Montenbruck, O., Gill, E., \& Lutze, F. 2002, Appl. Mech. Rev., 55, B27

\bibitem[{Morgado {et~al.}(2019)Morgado, Benedetti-Rossi, Gomes-J{\'u}nior,
  Assafin, Lainey, Vieira-Martins, Camargo, Braga-Ribas, Boufleur, Fabrega,
  {et~al.}}]{morgado2019}
Morgado, B., Benedetti-Rossi, G., Gomes-J{\'u}nior, A., {et~al.} 2019,
  Astronomy \& Astrophysics, 626, L4

\bibitem[{Morgado {et~al.}(2022)Morgado, Gomes-J{\'u}nior, Braga-Ribas,
  Vieira-Martins, Desmars, Lainey, D’aversa, Dunham, Moore, Bailli{\'e},
  {et~al.}}]{morgado2022}
Morgado, B., Gomes-J{\'u}nior, A., Braga-Ribas, F., {et~al.} 2022, The
  Astronomical Journal, 163, 240

\bibitem[{Park {et~al.}(2021)Park, Folkner, Williams, \& Boggs}]{park2021}
Park, R.~S., Folkner, W.~M., Williams, J.~G., \& Boggs, D.~H. 2021, The
  Astronomical Journal, 161, 105

\end{thebibliography}

\appendix

\setcounter{figure}{0}
\setcounter{table}{0}

\section{Consider parameter uncertainties}\label{appendix:considerParameters}

Table \ref{tab:apriori} provides the uncertainties for each of the consider parameters used in Juno state estimation. These uncertainties are based on mid-mission results \citep{iess2018,durante2020}.

\begin{table}[h]
        \small
        \caption{Uncertainties for the consider parameters.}
        \label{tab:apriori}
        \centering
        \begin{tabular}{l l }
                \hline
                \textbf{Parameter} & \textbf{Consider uncertainty}  \\ \hline
                $\mu$ & $8.9\times10^{9}$ [$\mathrm{m}^3\mathrm{s}^{-2}$]\\
                $J_2$ &  $1.7\times10^{-9}$ [-]\\
                $C_{21}$ & $2.3\times10^{-9}$ [-]\\
                $C_{22}$ & $1.1\times10^{-9}$ [-]\\
                $S_{21}$ & $1.5\times10^{-9}$ [-]\\
                $S_{22}$ & $1.0\times10^{-9}$ [-]\\
                $J_3$ & $3.3\times10^{-9}$ [-]\\
                $J_4$ & $2.4\times10^{-9}$ [-]\\
                $J_5$ & $4.2\times10^{-9}$ [-]\\
                $J_6$ & $6.7\times10^{-9}$ [-]\\
                $J_7$ & $1.2\times10^{-8}$ [-]\\
                $J_8$ & $2.1\times10^{-8}$ [-]\\
                $J_9$ & $3.6\times10^{-8}$ [-]\\
                $J_{10}$ & $6.5\times10^{-8}$ [-] \\
                $\alpha$ & $4.0\times10^{-5}$ [deg] \\
                $\delta$ & $5.0\times10^{-5}$ [deg] \\
                $\boldsymbol{a}_\mathrm{emp}$ & $2\times10^{-8}$ [$\mathrm{m}\mathrm{s}^{-2}$] \\ \hline
        \end{tabular}
        \tablefoot{Except for empirical accelerations $\boldsymbol{a}_\mathrm{emp}$, all parameters refer to Jupiter.}
\end{table}

\end{document}